# Magneto - transport characterization of Dy123 monodomain superconductors

M. Pekala
Department of Chemistry, University of Warsaw, Al. Zwirki i Wigury 101, PL-02-089 Warsaw, Poland

J. Mucha
Institute of Low Temperature and Structure Research, Polish Academy of Sciences, P.O. Box 1410, 50-950 Wroclaw, Poland

Ph. Vanderbemden
SUPRATECS, Institut d'Electricite Montefiore B28, University of Liege, B-4000 Liege, Belgium

R. Cloots
SUPRATECS, Department of Chemistry, B6 University of Liege, B-4000 Liege, Belgium

M. Ausloos
SUPRATECS, Institute of Physics, B5, University of Liege, B-4000 Liege, Belgium, Euroland

**Abstract**

The following report considers textured materials of the $DyBa_2Cu_3O_7$ type seeded with a Nd123 seed as initiator. They are grown with an excess 20% Dy211 phase on a $Dy_2O_3$ substrate. We report chemical characterizations, electrical resistivity, thermoelectric power and thermal conductivity over a broad temperature range as a function of an applied magnetic field up to 6 T. We show that specific features appear on the magneto thermal transport properties different in these materials from those found in single crystals and polycrystalline samples. We propose that two vortex regimes can be distinguished in the mixed phase, - due to the intrinsic microstructure. We calculate the viscosity, entropy and figure of merit of the samples.

# INTRODUCTION

The need for large single crystals of high critical temperature superconductors (HTSC) is often pointed out. They are necessary for studying fundamental properties. In applications, i.e. in order to avoid dissipation at grain boundaries, it is useful to have textured samples. There are different ways to do so. The texturing during growth difficulty is reduced by using, e.g. a magnetic field [1-6]. Two main advantages of the seed method are (i) that large monodomains can be grown up to a few centimeter wide and (ii) the orientation of the sample can be controlled by adjusting the orientation of the seed before the thermal cycle. In fact, disks and toroids have already been made through single or multi seeding techniques [3-6]. For deciding about the quality of a sample, beside comparing the microstructure of various samples one relies on measurements, like the (I,V) characteristics, the electrical resistance and the Meissner effect. It is well known that in the high-$T_C$ superconductor $YBa_2Cu_3O_{7-y}$, the $Y^{3+}$ ion can be replaced by most of the rare-earth ions, including $Dy^{3+}$ ion [7], without any appreciable effects on its superconducting properties. Some authors [8,9] are of the opinion that the partial substitution of Y3+ with a rare-earth ion can lead to changes in the crystal lattice which, in turn, can result in the formation of a stress field causing local flux pinning at the unit cell level.

Previous studies show, that 211 particles can lead to the mechanical reinforcement mechanism of 123 superconducting monodomains. It has been found through simulations and measurements that an optimal concentration of 211 particles, from the start is even useful for producing much 123 phase. For such fundamental and practical reasons, we have considered the production of bulk $DyBa_2Cu_3O_{7-x}$ superconducting ceramics through the seeding technique, in the presence of 211 excess [10]. Beside standard magnetic and electrical measurements we should like to emphasize that the thermal conductivity, k, the thermoelectric power or the Seebeck coefficient, s, and the Nernst-Ettinghausen coefficient, N, play an interesting role in the discussion of the phenomenon of high temperature superconductivity. Moreover, these magneto-thermal properties decide about the range of practical applications.

In the normal phase (T > $T_C$) they provide information on the electron scattering. In the superconducting phase (T < $T_C$), the tthermoelectric power vanishes more or less rapidly, whereas the thermal conductivity, k, and the Nernst-Ettinghausen coefficient, N , remain of a non-zero magnitude. In the case of k this is a manifestation of the transport of heat by phonons and normal electrons, while in the case of N it results from the movement of vortices under the influence of the temperature gradient. The simultaneous measurement of these three coefficients in the normal and superconducting phase guarantees that they are measured under the same thermodynamical conditions. This, in turn, provides a more reliable interpretation, since, e.g., a systematic measurement error is minimized and the behavior of one coefficient may suggest what physical mechanisms are responsible for the behavior of other ones.

A growth of large enough single crystals of high temperature superconductors is a cumbersome task. For studying the intrinsic properties of high temperature superconductors usually the textured polycrystalline samples are used in order to at least partially separate or limit an influence of the intergrain effects. In the case of systems belonging to the YBCO-123 class, the Dy123 superconductor is of interest because the magnetic moment borne by Dy atoms enables a very successful technique, i.e. magnetic texturing in external magnetic field [11-13] in order to produce the well aligned samples. Another technique recently used to grow large pieces of superconducting ceramics has been the seeding method. Highly ordered monodomain samples, containing a limited amount of defects, can be prepared by a classical isothermal melt-textured process.

This paper describes a new technological approach applied to prepare monodomain Dy123 samples as well as magneto —transport properties of such samples. The enhanced properties and the measured physical parameters are compared with those of the textured Dy123 samples prepared by the more classical method. According to our knowledge the effect of a magnetic field on the TEP of the dysprosium based superconductors is reported for the first time here.

**SAMPLE PREPARATION AND CHARACTERIZATION**

In the top-seeded melt-texturation method [3], one parallelepiped single crystal of e.g. $NdBa_2Cu_3O_{7-x}$ is often used as the seed material. This material has a much higher peritectic temperature $T_P = 1060\ ^0C$ than $DyBa_2Cu_3O_{7-x}$. Therefore at $1010\ ^0C$, Nd-123 does not decompose and constitutes the main nucleation center for $DyBa_2Cu_3O_{7-x}$ crystal growth. Since the lattice matching is favorable, the growing $DyBa_2Cu_3O_{7-x}$ phase is expected to conserve the crystallographic orientation of the seed. Moreover, the amount of crystal defects is expected to be minimized. The unreactive "seed method" has been shown to be able to produce large melt-textured monodomains with sizes up to a few centimeters [14].

Different seeds can be considered. Single crystals of $Dy_2O_3$ can be used for the production of large grains of Dy-123 superconducting materials [15]. $Dy_2O_3$ single crystals provide a sufficient concentration gradient of $Dy^{3+}$ in the liquid phase but cannot play the specific role of crystalline "seed" for epitaxy. The $Dy_2O_3$ seed "reacts" with the liquid phase and this reaction leads to the formation of a locally high concentration of small 211 particles, that will be easily dissolved and will favor the growth conditions of the 123 phase [15]. We have then suggested that a combination of Nd-123 and $Dy_2O_3$ seeds will provide one of the best solutions for producing large monodomains. However, $Dy_2O_3$ single crystals are very expensive materials. In order to reduce the cost of such reactive seeding techniques, we have investigated the seeding with polycrystalline $Dy_2O_3$ materials: both sintered and compacted powders. In the present case the $NdBa_2Cu_3O_{7-x}$ seed is on the top of the compressed pellet while a $Dy_2O_3$ polycrystal is used as substrate. The Dy123 + 20 weight % Dy211 pellet has been heat-treated at $1035\ ^0C$ for 2 hours and then cooled down to $980\ ^0C$ at $1\ ^0C/h$ cooling rate. The material has been reoxygenated for 4 days at $420\ ^0C$ under flowing oxygen.

An optical micrograph of the sample is given in Fig. 1. A $Dy_2O_3$ substrate was chosen in order to facilitate the growth of the 123 material by increasing locally the supersaturation. As previously reported [10, 16-18], a Dy211 layer, made from large 211 particles, can be observed at the interface between the 123 phase and the $Dy_2O_3$ bottom layer (Fig. 2). This high concentration of 211 particles creates the supersaturation level appropriate for inducing preferential growth of the 123 domain. The crystallographic orientation of the 123 material is uniform in throughout whole sample volume. By observing the twin patterns in a polarized light microscope, we can deduce that the crystallographic orientation of the micrographs (Fig. 3) shows the sample c-axis is nearly perpendicular ( – 5 degrees ) to the top surface of the $DyBa_2Cu_3O_{7-x}$ phase.

The microstructure shows many 211-free regions in the 123 domain created during its growth. Large bubbles are also created over the whole volume of the 211 "buffer" layer. A few cracks have been also created throughout volume of the sample. The average size of the 211 particles distributed in the 123 matrix is around 5 microns (Fig.4).

**MEASUREMENTS**

The magneto — transport measurements were performed for the relatively large monodomain with approximate dimensions: 4.5, 2 and 1.2 mm. Electric current was flowing in the ab - plane of the monodomains and the magnetic field was oriented along the c - axis. Conditions of measurement are practically the same as those described elsewhere [19]. Temperature stabilization and mean sweep rate were better than 0.01 K and less than 0.2 K/min, respectively.

**ELECTRICAL RESISTIVITY**

In the normal state the electrical resistivity of the system studied is of the order of 6 $\mu\Omega$m and rises almost proportionally with temperature up to about 14 $\mu\Omega$m at room temperature (Fig. 5 ). These parameters are comparable with the few published experimental data on electrical resistivity of these systems [11; 20-23]. The absolute values of electrical resistivity in the normal state are only 20 % higher as compared with the magnetic field textured Dy123 [11]. The difference may be due to the 20 % content of insulating Dy211 in the present sample. No magneto - resistance effects are found in the normal state resistivity. In zero external magnetic field the superconducting transition starts below 94 K and the transition interval width is about 1.5 K. When comparing with the magnetic field textured Dy123, the superconducting transition is enhanced by 4 K. Moreover the transition width interval is twice reduced. The onset temperature is almost unaffected by the magnetic field which remarkably broadens the transition interval up to 30 K at magnetic field of 6 T. Similar broadening of the transition interval was previously reported for Dy123+10 % of $BaZrO_3$ [20] but only for magnetic fields below 1 T.

The shape of the resistivity curve as a function of temperature in presence of a magnetic field is rather surprising and unusual. But it is very akin to that reported for melt processed $DyBa_2Cu_3O_7$ samples [24]. For single crystals the fall in resistivity systematically shifts to lower temperature, but the curve does not change shape. For polycrystalline samples, the drop in temperature usually presents some bending at some temperature below $T_C$, and a linear behavior due to the electron-defect contribution at the grain boundaries in a normal state. In the present case, there is no knee nor systematic shift. On the contrary there is a smooth curvature which is more and more pronounced as the field increases. However, one can observe a change in curvature toward a linear regime when the temperature is lowered. This crossover effect shifts from 87 K down to 65 K, when magnetic field raises from 2 to 6 T. This might indicate that the initial drop in resistivity is initially due to Josephson or tunnel junctions [25]. Finally electrical resistivity vanishes at the percolation temperature shifting from 89 K at zero magnetic field down to 61 K at 6 T field. A small suppression of transition temperature at zero field, as compared to the melt processed Dy123, may be related to the chemical imbalance at the $Dy_2O_3$ surface since the precursor material is drawn to $Dy_2O_3$ during a melting.

In the presence of a magnetic field the broadening of the transition can be reasonably understood within the concept of the conventional flux creep - flux flow theories. However one of the contributions seems to originate also in "giant" superconducting fluctuations that are due to the extremely short coherence length [26]. More generally the influence of magnetic field on the resistive transition may be accounted for by the integrated area below resistivity vs temperature curve [27-30] Fig. 6 shows that this area rises quickly with the field strength and may be fitted by the power law expression as $H^{1.74 \pm 0.03}$.

Moreover, the vortex viscosity calculated as $n(T) = \Phi_0 B_2 / \rho(T)$ is of the order of about $0.5 \times 10^{-9}$ Js/m$^3$ just below the transition temperature. The magnitude of the vortex viscosity is comparable to the values reported for the similar polycrystalline system $Y_{0.6}Dy_{0.4}Ba_2Cu_3O_{7-x}$ [31]. It is observed that the vortex viscosity rises with decreasing temperature as usually found for high temperature superconductors.

**THERMOELECTRIC POWER**

Absolute values of thermoelectric power (TEP) are relatively small in the normal state and not exceeding 1.2 µV/K in the temperature range studied. Such low TEP values are typical for the optimally doped dysprosium based superconductors as it was reported in [32]. In the normal state TEP decreases monotonically with raising temperature (Fig. 7) changing a sign around 200 K. A very similar thermoelectric behavior for the zero magnetic field was also reported for the analogous $(Nd_{1/3}Sm_{1/3}Gd_{1/3})Ba_2Cu_3O_7$ superconductors with 20 % addition of $Gd_2BaCuO_5$ micrometer size particles [33]. This superconductor exhibits the same magnitude of thermoelectric power and the negative slope of its temperature variation. The thermoelectric data for the Dy123 superconductor studied, show the superconducting onset temperature the same as that seen in electrical resistivity plots. A linear behavior is found between 120 and 280 K. According to our knowledge an effect of magnetic field onto the TEP is for a first time reported for the dysprosium based superconductors. It is observed that under a magnetic field the transition interval grows from 6 K at zero field up to more than thirty K at 6 T [34].

The normal state TEP remains unchanged by the applied magnetic field. The main variation occurs below 91 K or so. It is observed that under a magnetic field the transition interval grows from 6 K at zero field up to more than 30 K at 6 T. This corresponds to a shift of the percolation temperature from 88 K down to 56 K while magnetic field rises from zero to 6 T. The excess thermoelectric power in the mixed state defined as DQ(T) = Q(T,H) - Q(T,H=0) was derived from experimental data. The area below the DQ(T) curves was integrated and is plotted in Fig. 6 as a function of applied magnetic fields. A fitting procedure shows that the area of excess thermoelectric power raises with magnetic field as a power law with exponent 1.04 ± 0.02.

**NERNST EFFECT**

In the normal state above 100 K the Nernst signal ($Q_N = N B$) is negative and varies slowly with temperature, practically almost independent of the applied magnetic field strength [35]. When passing below the transition temperature the Nernst signal rises abruptly below 91 K and becomes positive. Taking into account residual voltages on the Nernst electrodes, the Nernst voltage generated by the magnetic field is determined as a difference between a signal registered at magnetic fields B > 0 and the signal corresponding to the zero field case. Even at magnetic field of 6 T the Nernst voltage is small and not exceeding $2 \times 10^{-7}$ V at maximum located at 84 K (Fig. 8). The same order of magnitude was reported for the magnetically textured Dy123 sample [11]. On further lowering the temperature the Nernst voltages diminishes gradually and vanishes below 51 K. A bump above the extrapolated line from the maximum in Nernst voltage is seen in the low temperature side of $T_C$ at nearly 65 K, - which is exactly the temperature in the resistivity curve separating the curved from the linear behavior. On the high temperature side it diminishes much faster and decays above 90 K. At lower magnetic fields from 1 to 4 T the Nernst voltage appears below 90 K and unfortunately is much noisy. However, the magnitude and the temperature interval of non-zero Nernst voltage shrinks quickly in weaker magnetic fields, as far as may be inferred from such noisy (not shown) plots.

The so called thermal Hall angle, $\alpha_{TH}$, between the directions of vortex velocity and temperature gradient is determined from the ratio of Nernst and Seebeck signals. Fig. 9 shows that $\alpha_{TH}$ equals zero above 90 K. Below 90 K values of $\alpha_{TH}$ rise abruptly with bump at 84 K, when the Nernst effect has its maximum. The rise is understood as due to the vortex velocity gradually deviating from a direction of temperature gradient. The maximum value of $\alpha_{TH} = 88$ degrees is achieved at 77 K. A bump again appears at 65 K - corresponding to the crossover from the curved to linear resistivity variation. The entropy carried by vortices equal to $S(T) = \Phi_0 N / \rho(T)$, is relatively low as could be expected for a superconductor exhibiting a Nernst effect of low magnitude. The transport entropy increases below 90 K when material enters the mixed state and attains the local maximum at 84 K followed by a bump at 65 K (Fig. 10). These features are similar to those found for Nernst signal and vortex viscosity too.

**THERMAL CONDUCTIVITY**

Fig. 11 shows a temperature variation of thermal conductivity, k(T), in the magnetic fields varying from zero to 6 T. In the normal state the thermal conductivity diminishes monotonically with a decreasing temperature from the room temperature value of 7.3 W/Km. Such a relatively high value of thermal conductivity, being comparable with those for single crystals [37], confirms a high quality of the sample studied. In this temperature range the phonons are scattered by electrons and large linear defects. This explains the curvature in the data. On decreasing temperature below the superconducting transition of 90 K the thermal conductivity starts to increase and achieves a maximum of 6.6 W/Km about 62 K. This maximum occurring due to the reduced phonon - electron scattering in superconducting state, is characteristic for the high temperature superconductors with thermal conductivity dominated by the phonon contribution [19, 36-40]. Applying the Wiedemann - Franz relation one may estimate that the electronic contribution to thermal conductivity does not exceeds 7 % in the normal state. This confirms that heat is overwhelmingly transported by phonons. Between 62 and 90 K the thermal conductivity in zero magnetic field fits perfectly to a $T^{-1}$ variation being characteristic for the dominant anharmonic phonon - phonon scattering. The temperature dependence of thermal conductivity on the low temperature side of the maximum may be fitted to the $k \propto T$ relation, which reveals the elastic scattering of phonons on point defects. When applying a magnetic field, the normal state thermal conductivity is suppressed by a few percent. The thermal conductivity maximum, well pronounced for zero magnetic field, is gradually suppressed and shifted to lower temperatures by fields of 1 and 2 T. This proves that in the magnetic field the electron —phonon scattering survives down to lower temperatures. The maximum is then washed out by the 6 T magnetic field (Fig. 11). Thermal conductivity falls down quickly below about 50 K, in a range where phonon scattering on defects and chemical impurities play a role.

The figure of merit equal to $Z = Q^2 / k \rho$ shows that the cooling efficiency $ZT$ of the material studied is of the order of the $2 \times 10^{-5}$, which makes a material suitable as a passive branch for Peltier cooling devices. Application of magnetic field up to 8 T does not affect the $ZT$ value noticeably.

**CONCLUSION**

The search for new HTSC with high critical temperatures and the ability to carry an electric current of high density $J_C$ is surely of interest. However to develop and optimize chemical compositions and treatment conditions leading to materials with technologically required physical, chemical and technical characteristics / properties requires technology to be supported toward fundamental science. Therefore various technological treatments used to confine the weak links limiting the current density are useful routes of research. The above report on textured seeded $DyBa_2Cu_3O_7$ materials is presented in such a framework. We have found that the relatively high electrical and thermal conductivities of textured seeded Dy123 monodomains prove that this method enables the high quality materials. Such monodomains are highly structurally ordered. Thus, the intergrain connectivity is improved. It was also checked that a magnetic field of 6 T does not dramatically suppress the percolation temperatures as determined from electrical resistivity and thermoelectric power (Fig. 12). The relatively small differences between the temperatures detected by electrical resistivity and thermoelectric power may be due to a precision on temperature measurement in a sample with gradient of T and without gradient of T, as seen by many along time ago and in particular in our group [2,6]. Values of the figure of merit show a material studied is suitable for a passive branch of Peltier cooling devices.

**Acknowledgement.** Work supported in parts by KBN grant 7T08A 02820 and Polish - Belgian Exchange Program.

**FIGURE CAPTIONS**

Fig. 1. Optical micrograph of the DyBa$_2$Cu$_3$O$_{7-x}$ monodomain (left upper corner) grown on a Dy$_2$O$_3$ substrate (right bottom corner). Letters a and b point to areas enlarged in next micrographs.

Fig. 2. The Dy211 interlayer observed at the interface between the Dy123 phase and the Dy$_2$O$_3$ bottom substrate layer.

Fig. 3. The crystallographic orientation of the Dy123 material is uniform as shown by the c-axis nearly perpendicular ( – 5 degrees ) to the top surface of the DyBa$_2$Cu$_3$O$_{7-x}$ phase.

Fig. 4. A fine structure of cracks and a distribution of the Dy211 particles in the Dy123 matrix in area c .

Fig. 5. Electrical resistivity normalized to R$_{100K}$ vs temperature for various magnetic fields.

Fig. 6. Magnetic field dependence of the area below the electrical resistivity and thermoelectric power vs. temperature curves.

Fig. 7. Temperature variation of thermoelectric power measured at various magnetic fields.

Fig. 8. Temperature variation of Nermst voltage measured at 6 T magnetic field.

Fig. 9. Temperature variation of thermal Hall angle in 6 T magnetic field.

Fig. 10. Temperature variation of transport entropy in 6 T magnetic field.

Fig. 11. Temperature variation of thermal conductivity measured at various magnetic fields.

Fig. 12. Percolation lines determined from electrical resistivity and thermoelectric power, as a temperature where electrical resistivity and thermoelectric power, respectively, vanish in magnetic field

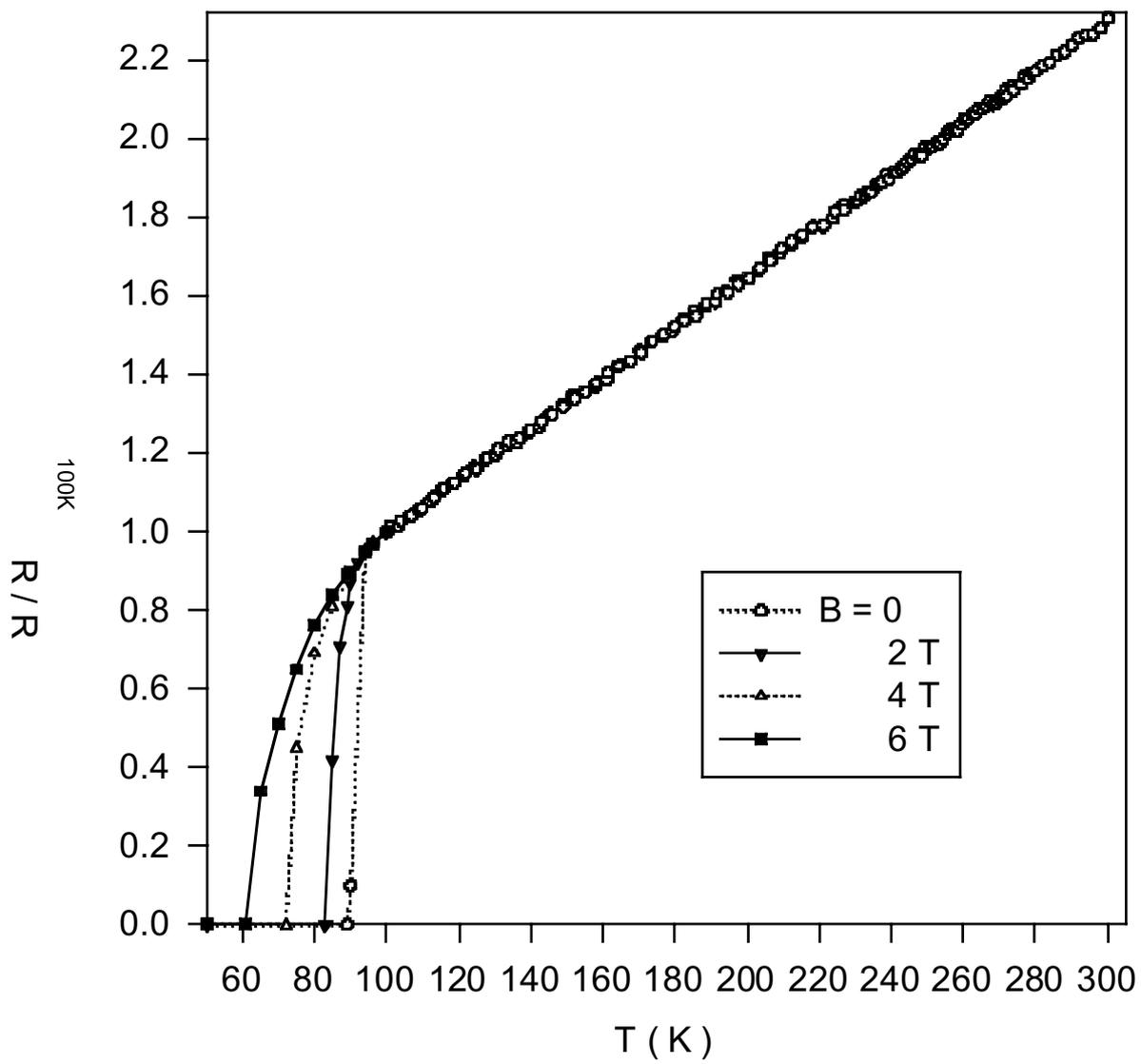

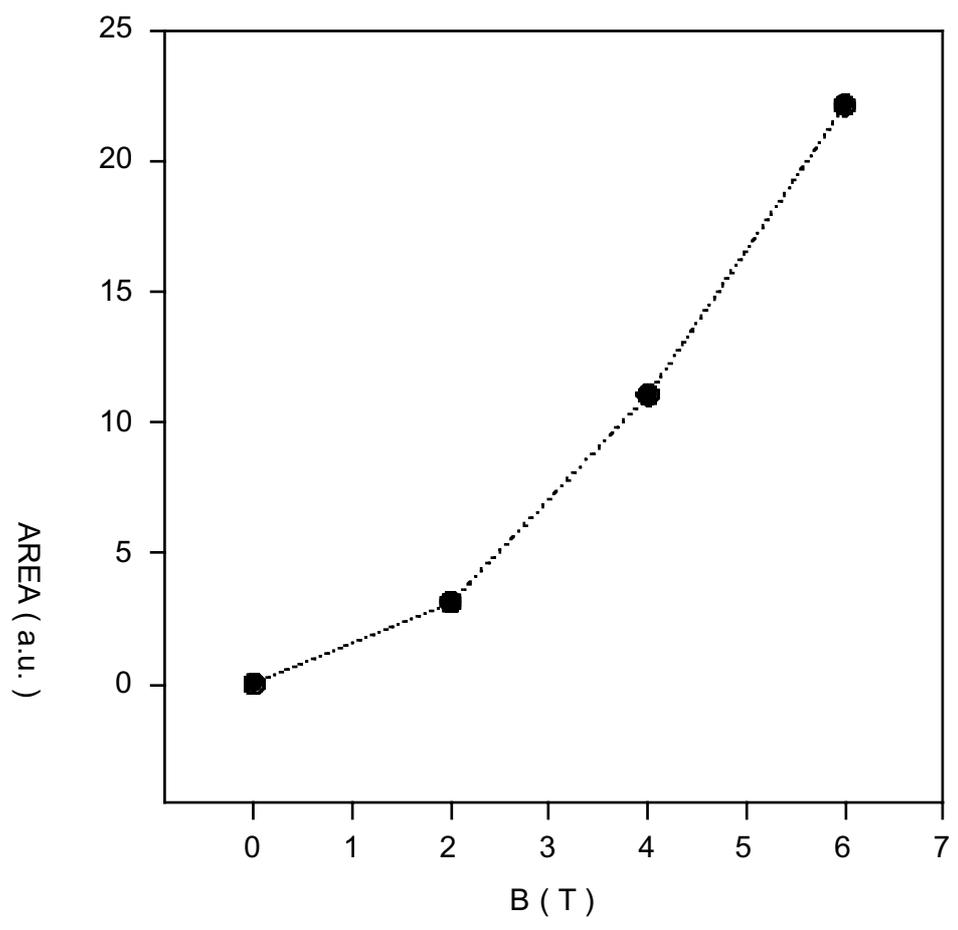

Fig. 7

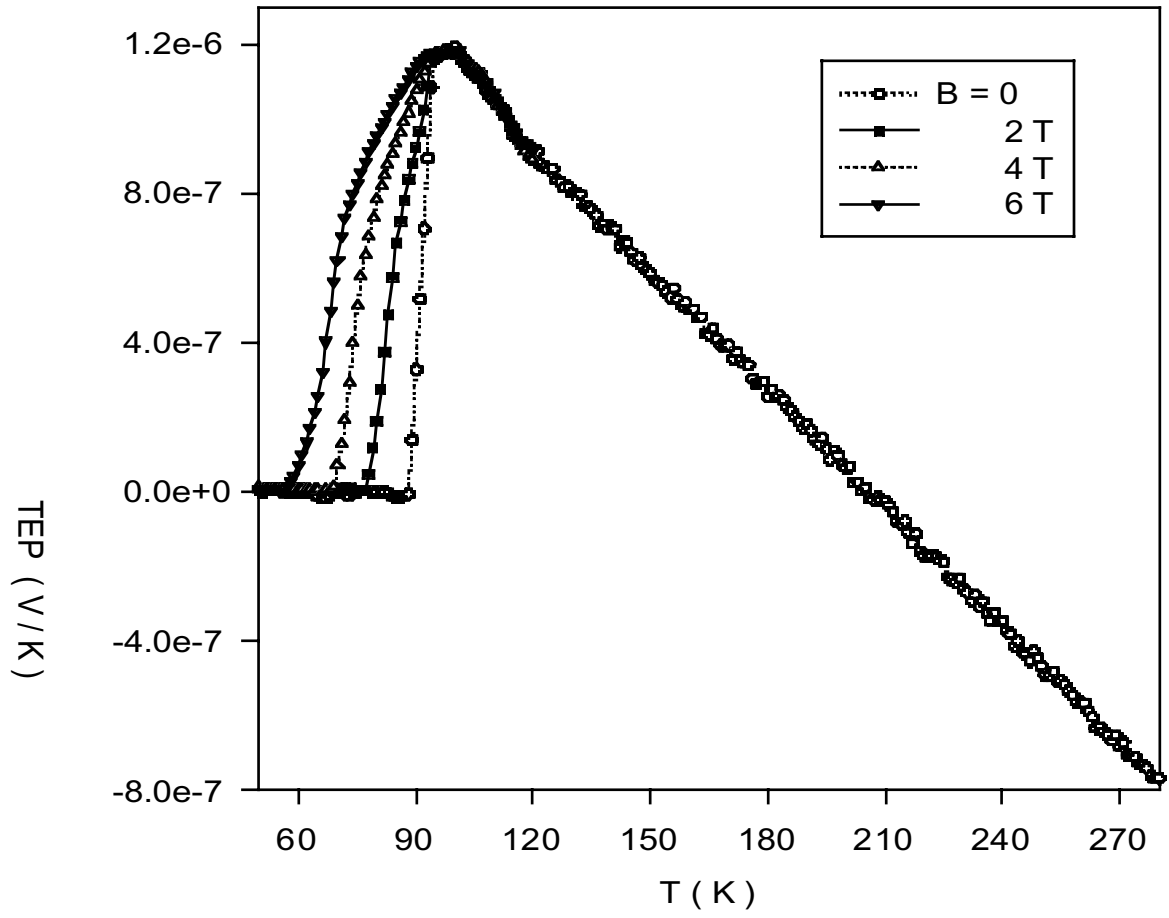

Fig. 8

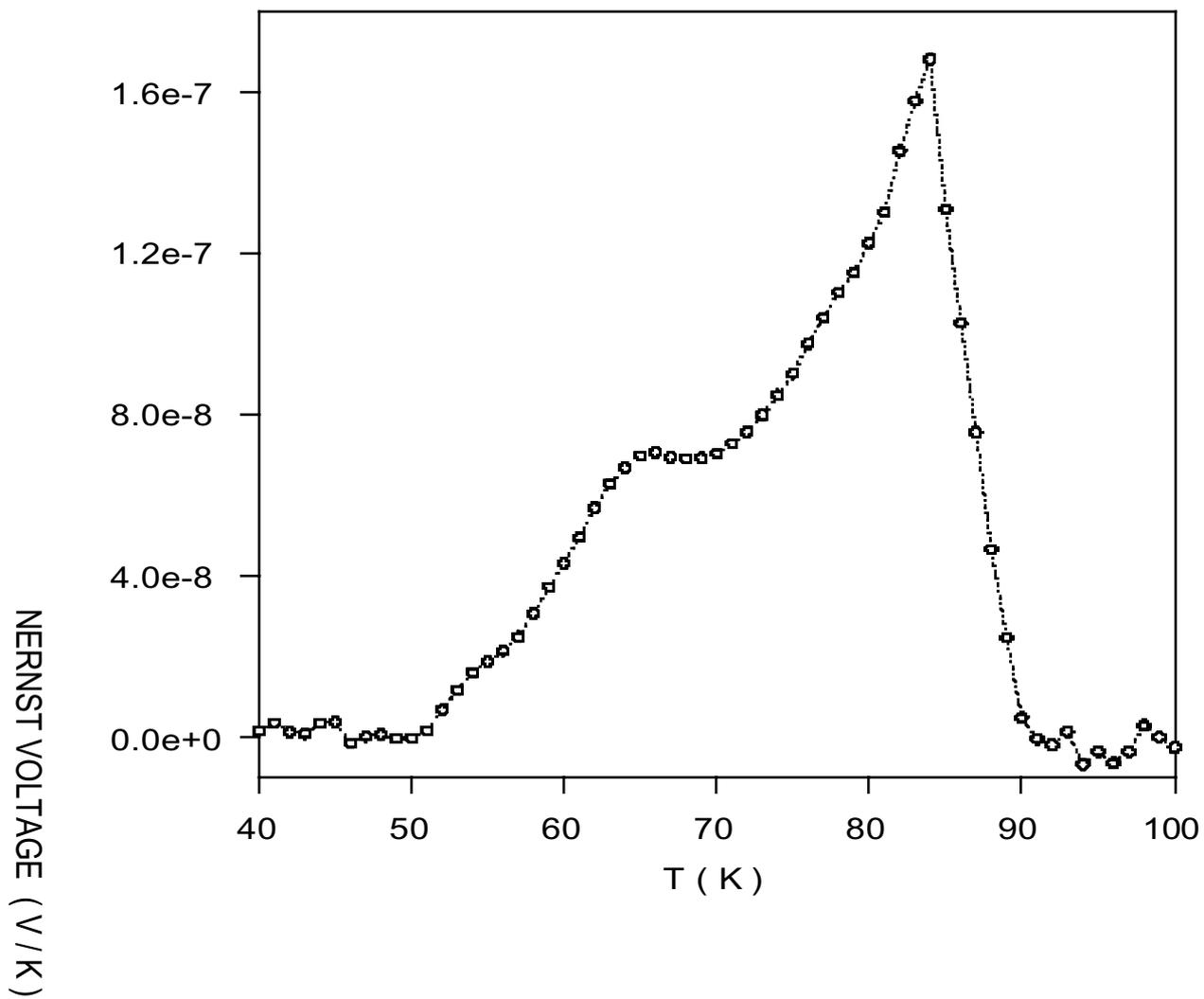

Fig. 9

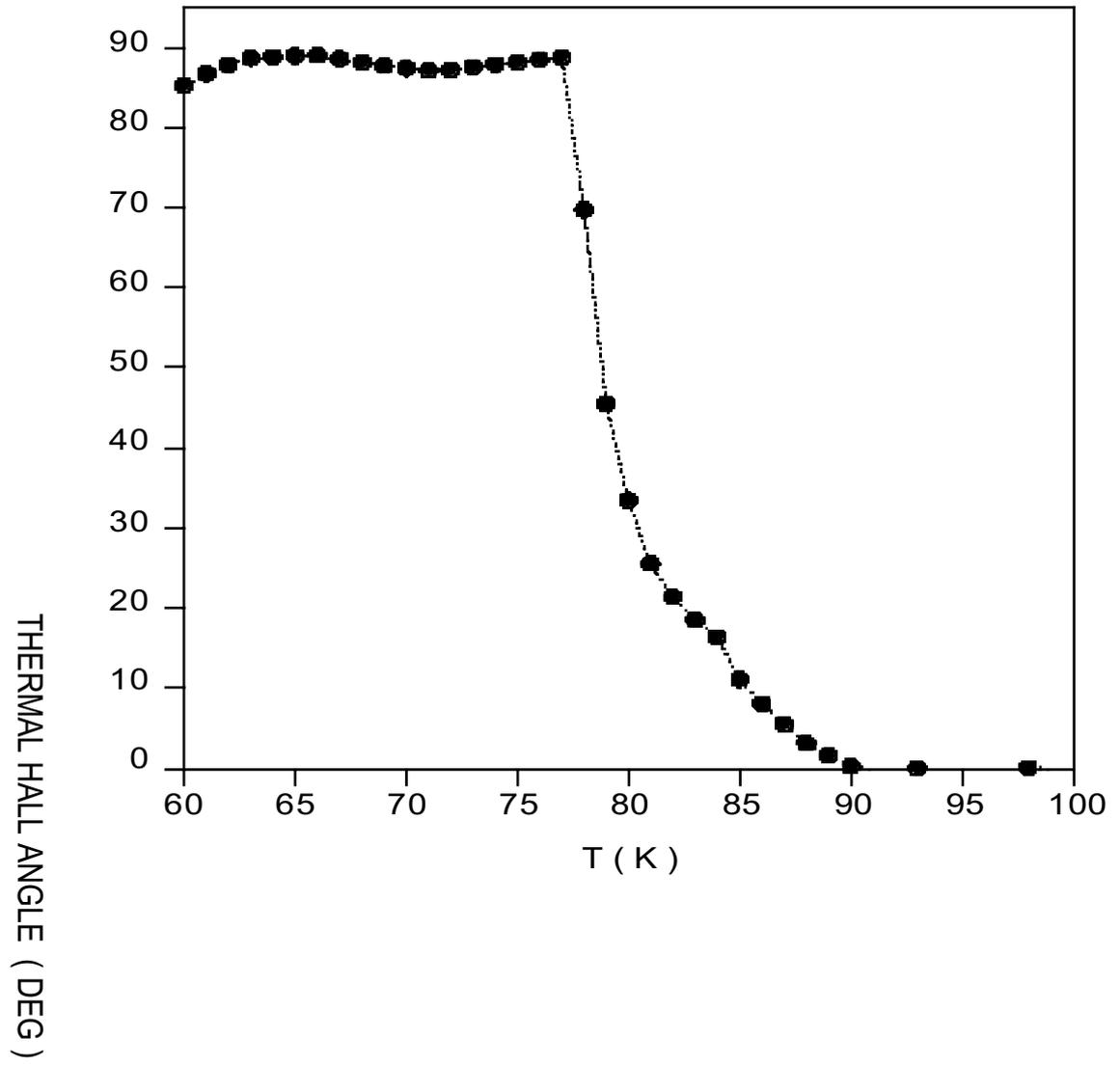

Fig. 10

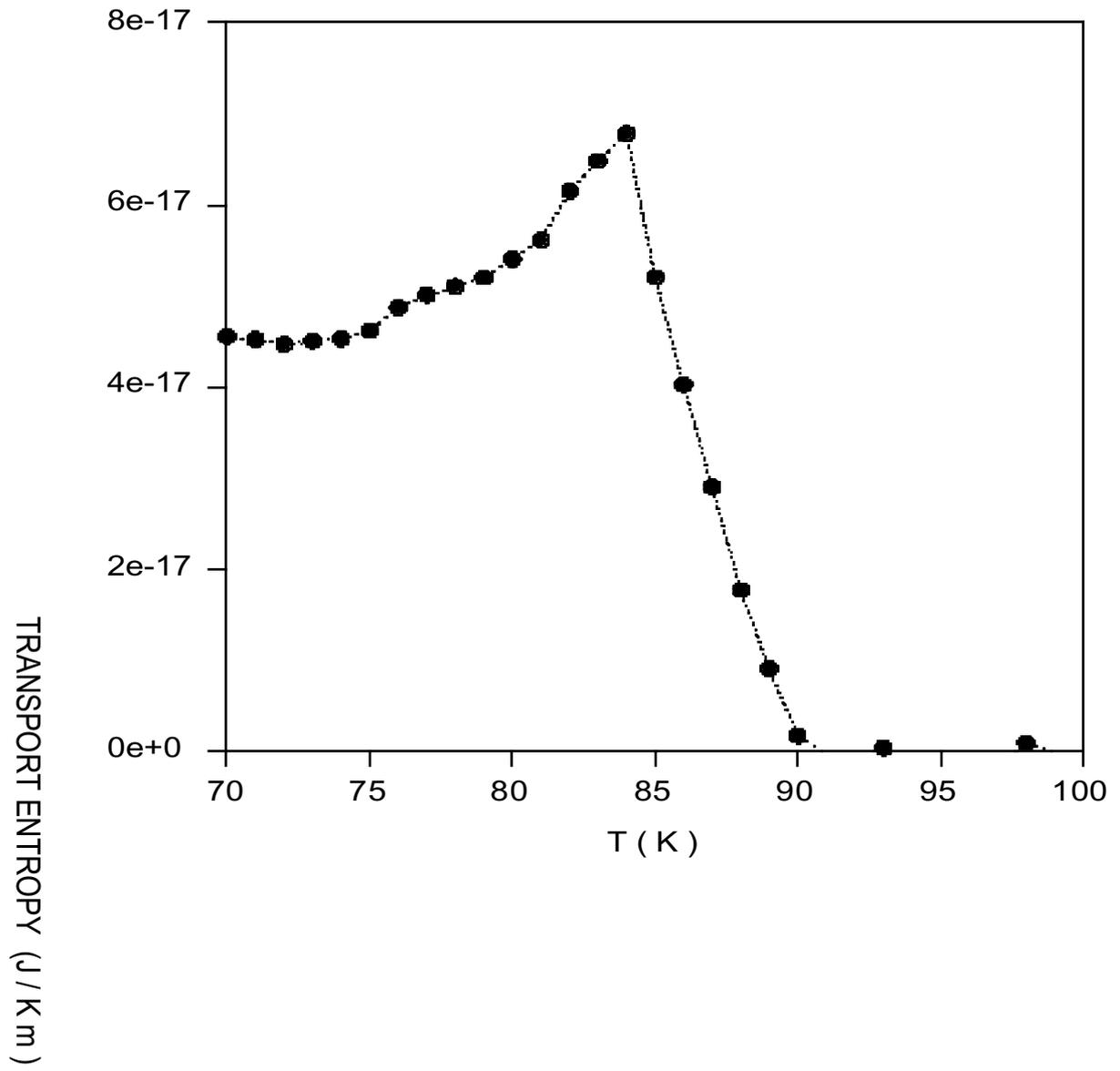

Fig. 11

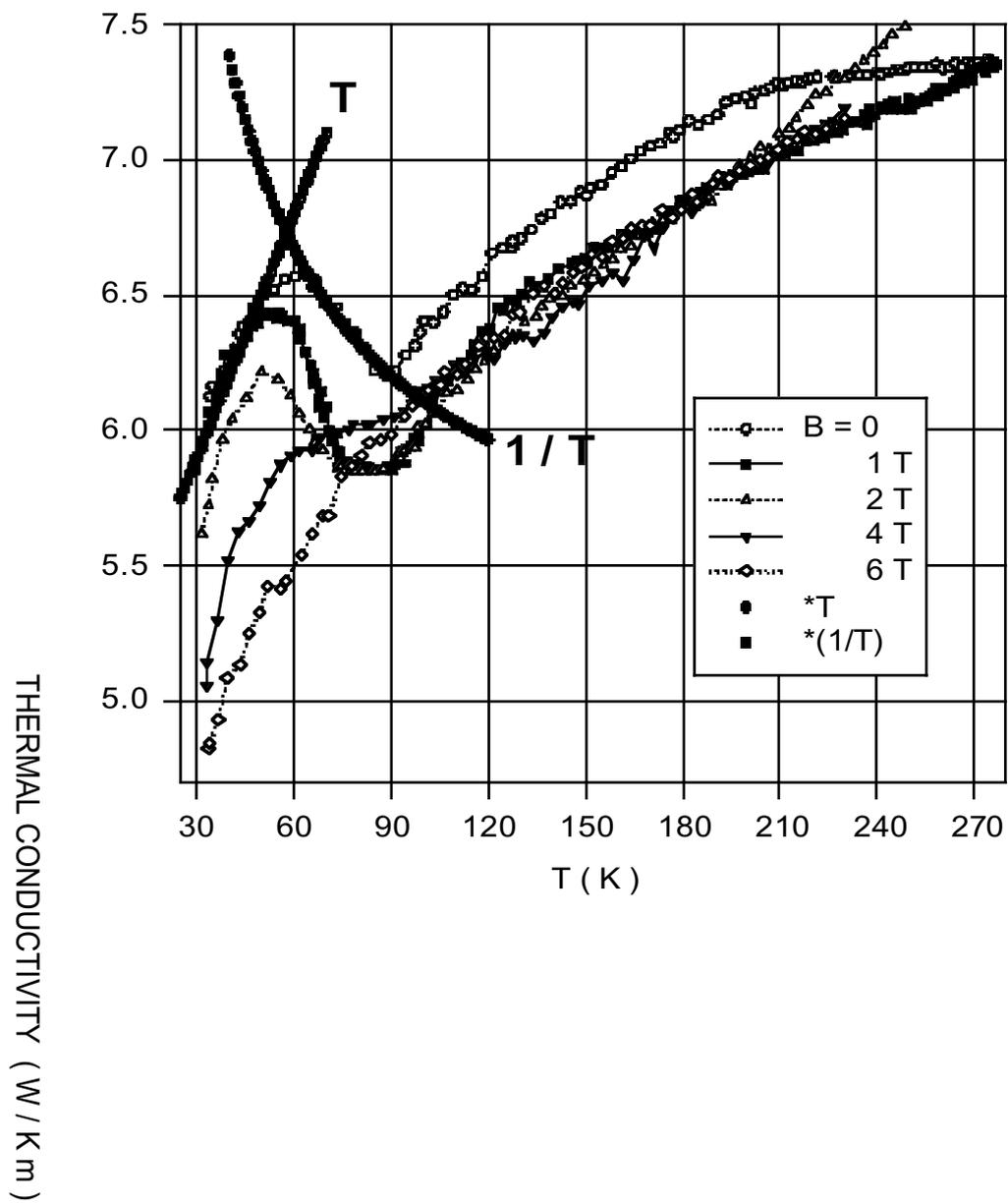

Fig. 12

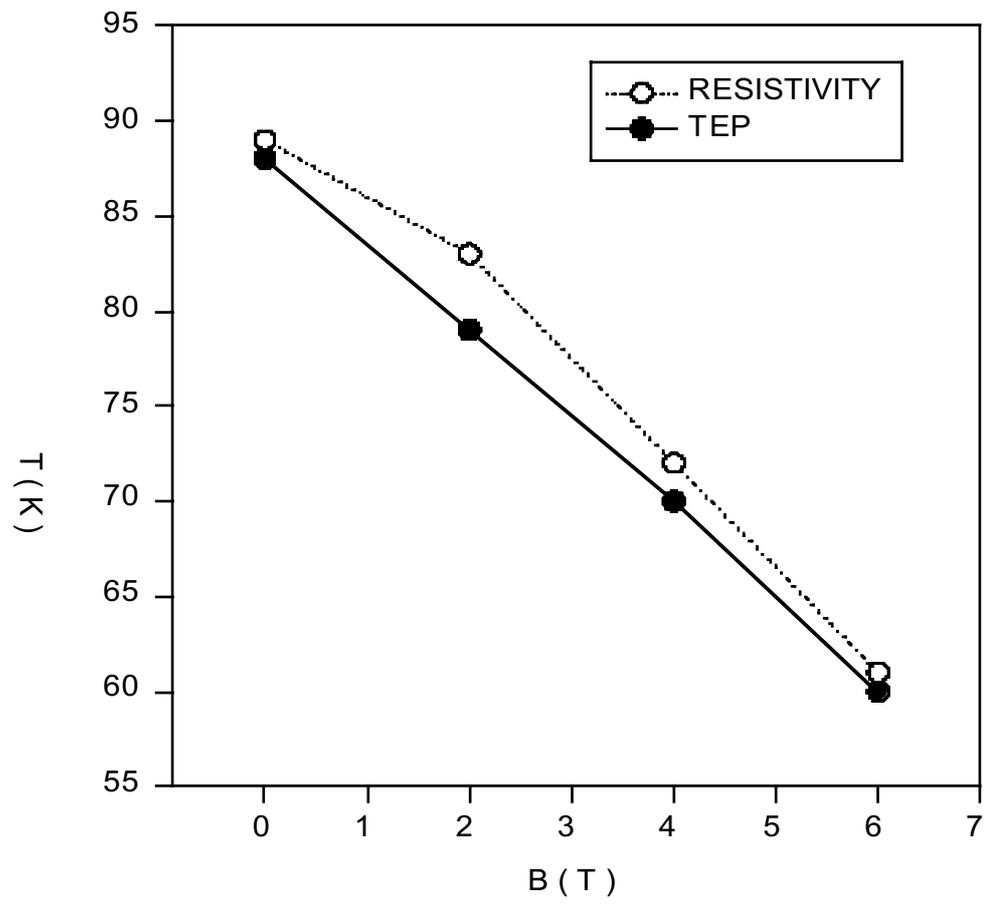